\documentclass[letterpaper,english,reprint,aps,prb]{revtex4-1}
\usepackage[T1]{fontenc}
\usepackage[latin9]{inputenc}
\usepackage{amsmath}
\usepackage{amssymb}
\usepackage{graphicx}

\makeatletter

\@ifundefined{showcaptionsetup}{}{%
 \PassOptionsToPackage{caption=false}{subfig}}
\usepackage{subfig}
\makeatother

\usepackage{babel}
\begin{document}

\preprint{APS/123-QED}

\title{Decoupling between first sound and second sound \\
 in dilute \textsuperscript{3}He -- superfluid \textsuperscript{4}He
mixtures}

\author{T. S. Riekki}

\email{tapio.riekki@aalto.fi}

\author{M. S. Manninen}

\author{J. T. Tuoriniemi}

\affiliation{Low Temperature Laboratory, Department of Applied Physics, Aalto
University, P.O. BOX 15100 FI-00076 AALTO }

\date{\today}
\begin{abstract}
Bulk superfluid helium supports two sound modes: first sound is an
ordinary pressure wave, while second sound is a temperature wave,
unique to inviscid superfluid systems. These sound modes do not usually
exist independently, but rather variations in pressure are accompanied
by variations in temperature, and vice versa. We studied the coupling
between first and second sound in dilute \textsuperscript{3}He --
superfluid \textsuperscript{4}He mixtures, between $1.6\,\mathrm{K}$
and $2.2\,\mathrm{K}$, at \textsuperscript{3}He concentrations ranging
from $0$ to $11\,\%$, under saturated vapor pressure, using a quartz
tuning fork oscillator. Second sound coupled to first sound can create
anomalies in the resonance response of the fork, which disappear only
at very specific temperatures and concentrations, where two terms
governing the coupling cancel each other, and second sound and first
sound become decoupled.
\begin{description}
\item [{PACS~numbers}] 67.60.-g, 67.25.dt{\small \par}
\end{description}
\end{abstract}



\maketitle

\section{\label{sec:Introduction}Introduction}

There exist two possible sound modes in bulk superfluid helium: first
sound is an ordinary pressure (or density) wave, whereas second sound
is a temperature (or entropy) wave. Second sound is unique to superfluid
systems, where temperature fluctuations can propagate as waves due
to the existence of two independent velocity fields of normal fluid
and superfluid component. In normal systems all temperature fluctuations
are so strongly damped that such a wave cannot exist. In terms of
Tisza's \cite{Tisza1,Tisza2} and Landau's \cite{Landau} two-fluid
model for superfluid helium, first sound is the mode where normal
fluid and superfluid component oscillate in phase, while in second
sound they oscillate antiphase. Since \textsuperscript{3}He in dilute
\textsuperscript{3}He --\textsuperscript{4}He mixtures is in normal
state, it flows with the normal fluid component, which gives another
interpretation for second sound: \textsuperscript{3}He concentration
wave. 

First sound and second sound do not usually exist independent from
each other, but rather pressure fluctuations of first sound are accompanied
by second sound's fluctuations in temperature, and vice versa. In
pure \textsuperscript{4}He, the coupling is due to the thermal expansion
of the liquid, even though it is extremely small. The addition of
\textsuperscript{3}He modifies, not only the superfluid transition
temperature of \textsuperscript{4}He, but also the coupling between
the two sound modes. In this paper, we show that it is possible to
find conditions where second sound and first sound become decoupled
from each other, when thermal expansion contribution and \textsuperscript{3}He
contribution to the coupling cancel each other.

Our studies were conducted using quartz tuning forks, which are commercially
mass produced piezoelectric oscillators, whose intended frequency
is usually around $32\,\mathrm{kHz}$. They can be used to measure,
for example, temperature, pressure, concentration, viscosity, and
turbulence in liquid helium \cite{Blaauwgeers,Pentti_Osmotic_pressure,Pentti_etal_solubility,Bradley_Transition2turbulence}.
Velocity of second sound in superfluid helium is of order $10\,\mathrm{m/s}$,
and its characteristic wavelength, at the used frequency, matches
the dimensions of common quartz tuning forks. Consequently, at certain
temperatures, second sound is able to form standing waves in the fluid
surrounding the fork, whereas first sound, with velocity of order
$100\,\mathrm{m/s}$, is usually not. When the sound modes are coupled,
second sound can drive first sound, and the effect of this driven
first sound can be seen as an anomaly in the resonance response of
the fork.

These kind of anomalies in the quartz tuning fork response, or second
sound resonances, have been observed before \cite{Pentti_Rysti_Salmela,Salmela_Tuoriniemi_Rysti,Salmela_fixedPoints},
but the detailed mechanism producing these anomalies has not been
thoroughly investigated. Calculations of the coupling factors between
first and second sound in helium mixtures have been presented before
by Brusov \emph{et al.} \cite{Brusov}, but, as first noticed by Rysti
\cite{Juho}, they made a sign error in their calculations, which
prevented them from noticing the decoupling behavior.

Before presenting the results of our experiment, we first go briefly
over the revised calculation of the coupling factors governing the
conversion between first sound and second sound.

\section{Sound conversion\label{sec:Sound-Conversion}}

Coupling between first sound and second sound in pure \textsuperscript{4}He
is due to the thermal expansion of the liquid which connects changes
in temperature to changes in pressure and vice versa. Since the thermal
expansion coefficient of superfluid \textsuperscript{4}He is extremely
small, the coupling between the sound modes is very weak. The addition
of the lighter isotope \textsuperscript{3}He modifies the coupling
so that at low concentrations the coupling becomes even weaker, eventually
vanishing at specific temperatures and concentrations. As the concentration
is further increased, the \textsuperscript{3}He contribution to the
coupling starts to dominate the system and the coupling grows stronger.

We can obtain expressions for sound conversions by starting from the
linearized two-fluid hydrodynamical equations presented by Khalatnikov
\cite{Khalatnikov_TheoryofSuperfluidity}, from which we reach a set
of equations characterizing the sound propagation in \textsuperscript{3}He
--\textsuperscript{4}He mixtures
\begin{equation}
\frac{\partial^{2}\rho}{\partial t^{2}}=\nabla^{2}P,\label{eq:1st_D}
\end{equation}
\begin{equation}
\frac{\rho_{\mathrm{n}}}{\rho_{\mathrm{s}}\sigma}\frac{\partial^{2}\sigma}{\partial t^{2}}=\sigma\nabla^{2}T+c\nabla^{2}\left(\frac{Z}{\rho}\right),\:\mathrm{and}\label{eq:2nd_D}
\end{equation}
\begin{equation}
\frac{1}{\sigma}\frac{\partial\sigma}{\partial t}=\frac{1}{c}\frac{\partial c}{\partial t}.\label{eq:3rd_D}
\end{equation}
Here $\rho$, $t$, $P$, $\sigma$, $T$, and $c$ are density, time,
pressure, specific entropy, temperature, and \textsuperscript{3}He
mass concentration, respectively, whereas $\rho_{\mathrm{n}}$($\rho_{\mathrm{s}}$)
is normal fluid (superfluid) density. Furthermore, $Z\equiv\rho\left(\mu_{3}-\mu_{4}\right)$,
where $\mu_{3}$ and $\mu_{4}$ are the chemical potentials of \textsuperscript{3}He
and \textsuperscript{4}He, respectively. Eq. \eqref{eq:1st_D} is
the first sound wave equation, Eq. \eqref{eq:2nd_D} the second sound
wave equation, and Eq. \eqref{eq:3rd_D} is the result of the conservation
of entropy and \textsuperscript{3}He ``impurities''.

Next, we can choose $T$, $P$, and $c$ to be our independent variables,
and consider small perturbations around their equilibrium values,
so that $T=T_{0}+\widetilde{T}\left(\mathbf{r},t\right)$, where $T_{0}$
is the equilibrium value, and $\widetilde{T}$ the small deviation,
and similarly for the other variables. We further assume that the
perturbations are of plane wave form $\propto\exp\left(i\omega\left(\frac{z}{u}-t\right)\right)$,
where $\omega$ is the angular frequency, $u$ the velocity of the
wave, and $z$ the direction of propagation. When we next eliminate
$c$ from eqs. \eqref{eq:1st_D}-\eqref{eq:3rd_D} we obtain a linear
set of equations of form
\begin{align}
A_{00}(u^{2})\widetilde{T}+A_{01}(u^{2})\widetilde{P} & =0\label{eq:1st_L}\\
A_{10}(u^{2})\widetilde{T}+A_{11}(u^{2})\widetilde{P} & =0\label{eq:2nd_L}
\end{align}
where Eq. \eqref{eq:1st_L} is the linearized first sound wave equation,
and Eq. \eqref{eq:2nd_L} the linearized second sound wave equation.
If we assume that the eigenvalues of this system are pure first sound
($u_{1}$) and pure second sound ($u_{2}$), we can consider the two
equations above independently. In order to see how second sound creates
first sound, we insert $u=u_{2}$ in the first sound wave equation
\eqref{eq:1st_L} \emph{i.e.} we use second sound as source for the
first sound. This way, with appropriate simplifications, we get
\begin{align}
\widetilde{P} & =-\frac{A_{00}(u_{2}^{2})}{A_{01}(u_{2}^{2})}\widetilde{T}\nonumber \\
 & =\left[\left(\frac{\partial\rho}{\partial T}\right)_{P,c}+\frac{c_{0}}{\bar{\sigma}}\left(\frac{\partial\rho}{\partial c}\right)_{T,P}\left(\frac{\partial\sigma}{\partial T}\right)_{P,c}\right]\frac{u_{1}^{2}u_{2}^{2}}{u_{1}^{2}-u_{2}^{2}}\widetilde{T}\nonumber \\
 & \equiv\alpha\widetilde{T},\label{eq:2 to 1}
\end{align}
where $\bar{\sigma}\equiv\sigma_{0}-c_{0}\frac{\partial\sigma}{\partial c}$,
and $\alpha$ is the coupling factor governing conversion of second
sound into first sound. Similarly, if we insert $u=u_{1}$ in the
second sound wave \linebreak equation \eqref{eq:2nd_L}, we obtain
\begin{align}
\widetilde{T} & =-\frac{A_{11}(u_{1}^{2})}{A_{10}(u_{1}^{2})}\widetilde{P}=\frac{U_{1}^{2}\left(\frac{\partial\rho}{\partial T}\right)_{P,c}+c_{0}\bar{\sigma}\left(\frac{\partial\rho}{\partial c}\right)_{T,P}}{\rho_{0}^{2}\bar{\sigma}^{2}-\rho_{0}^{2}U_{1}^{2}\left(\frac{\partial\sigma}{\partial T}\right)_{P,c}}\widetilde{P}\nonumber \\
 & \equiv\beta\widetilde{P,}\label{eq:1 to 2}
\end{align}
where $U_{1}^{2}\equiv u_{1}^{2}\frac{\rho_{\mathrm{n}}}{\rho_{\mathrm{s}}}-c_{0}^{2}\left(\frac{\partial\left(Z/\rho\right)}{\partial c}\right)_{T,P}$,
and $\beta$ is the coupling factor characterizing conversion of first
sound into second sound. These equations are similar to what Brusov
\emph{et al. }\cite{Brusov}\emph{ }had obtained, except that, as
noted by Rysti \cite{Juho}, they made a sign error in the bracketed
term of Eq. \eqref{eq:2 to 1}, which prevented them from noticing
the possibility of decoupling between the two sound modes. For superfluid
helium $\left(\partial\rho/\partial T\right)_{P,c}$, which is proportional
to the thermal expansion coefficient, is positive, and since $c_{0}/\bar{\sigma}$,
and $\left(\partial\sigma/\partial T\right)_{P,c}$ are also positive,
but $\left(\partial\rho/\partial c\right)_{T,P}$ is negative, it
is possible that the two terms in brackets cancel out each other at
certain \textsuperscript{3}He concentrations and temperatures, resulting
in first sound decoupling from second sound. The velocity term of
$\alpha$ does not change sign since $u_{1}>u_{2}$ always. On the
other hand, the coupling factor $\beta$ is small, always negative,
and practically constant in the temperature and concentration region
of our experiment. Only very close to \textsuperscript{4}He superfluid
transition temperature, $T_{\lambda}$, its value starts to depart
from the constant value. Since $\beta$ is always finite, first sound
can always generate second sound in the region considered here, whereas
there exists specific temperatures and concentrations where the opposite
is not possible; second sound cannot always create first sound.

\begin{figure}
\includegraphics[width=9cm]{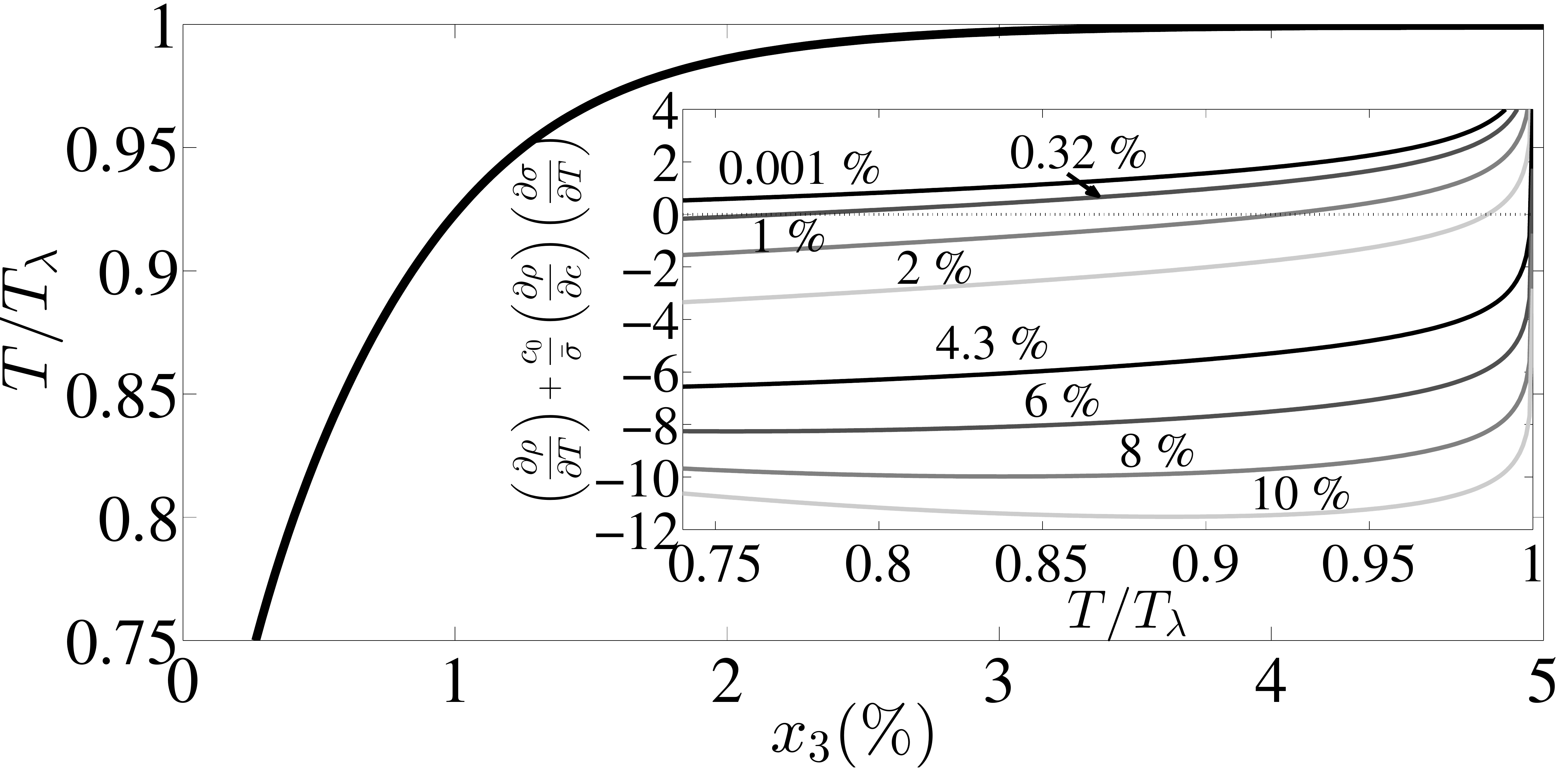}

\caption{Calculated temperature relative to $^{4}\mathrm{He}$ superfluid transition
temperature, $T_{\lambda}=T_{\lambda}(x_{3})$, where first sound
decouples from second sound ($\alpha=0$), as a function of molar
$^{3}\mathrm{He}$ concentration, $x_{3}$. \emph{Inset} shows the
values of the bracketed term of $\alpha$ in Eq. \eqref{eq:2 to 1}
as a function of relative temperature at different $^{3}\mathrm{He}$
concentrations.\label{fig:decoupling_calc}}
\end{figure}

Fig. \ref{fig:decoupling_calc} shows the decoupling conditions, as
well as calculated values for the bracketed term of the coupling factor
$\alpha$, obtained by using the entropy data of Refs. [\onlinecite{Roberts_Sherman,Kramers_Wasscher_Gorter,Hill_Lounasmaa}],
and by evaluating the superfluid and normal fluid densities in mixture
according to Refs. [\onlinecite{Juho,Wilks_LiquidSolid,Baym_Pethick,Matti,Donnelly}].
Furthermore, the density of the mixture was evaluated by using molar
volume formula given by Dobbs \cite{Dobbs}, and the pure \textsuperscript{4}He
density formula given by Niemela and Donnelly \cite{Niemela_Donnelly}
scaled by the \textsuperscript{3}He concentration dependence of $T_{\lambda}$,
to produce correct density behavior near $T_{\lambda}$. 

At $0.75T_{\lambda}\approx1.6\,\textrm{K}$, the decoupling occurs
already around $0.3\,\%$ molar \textsuperscript{3}He concentration,
and at higher temperatures the decoupling condition moves to higher
concentrations, up to about $3\,\mathrm{\%}$ near $T_{\lambda}$.

\section{Experimental details\label{sec:Experimental-Details}}

\begin{figure}
\includegraphics[width=7cm]{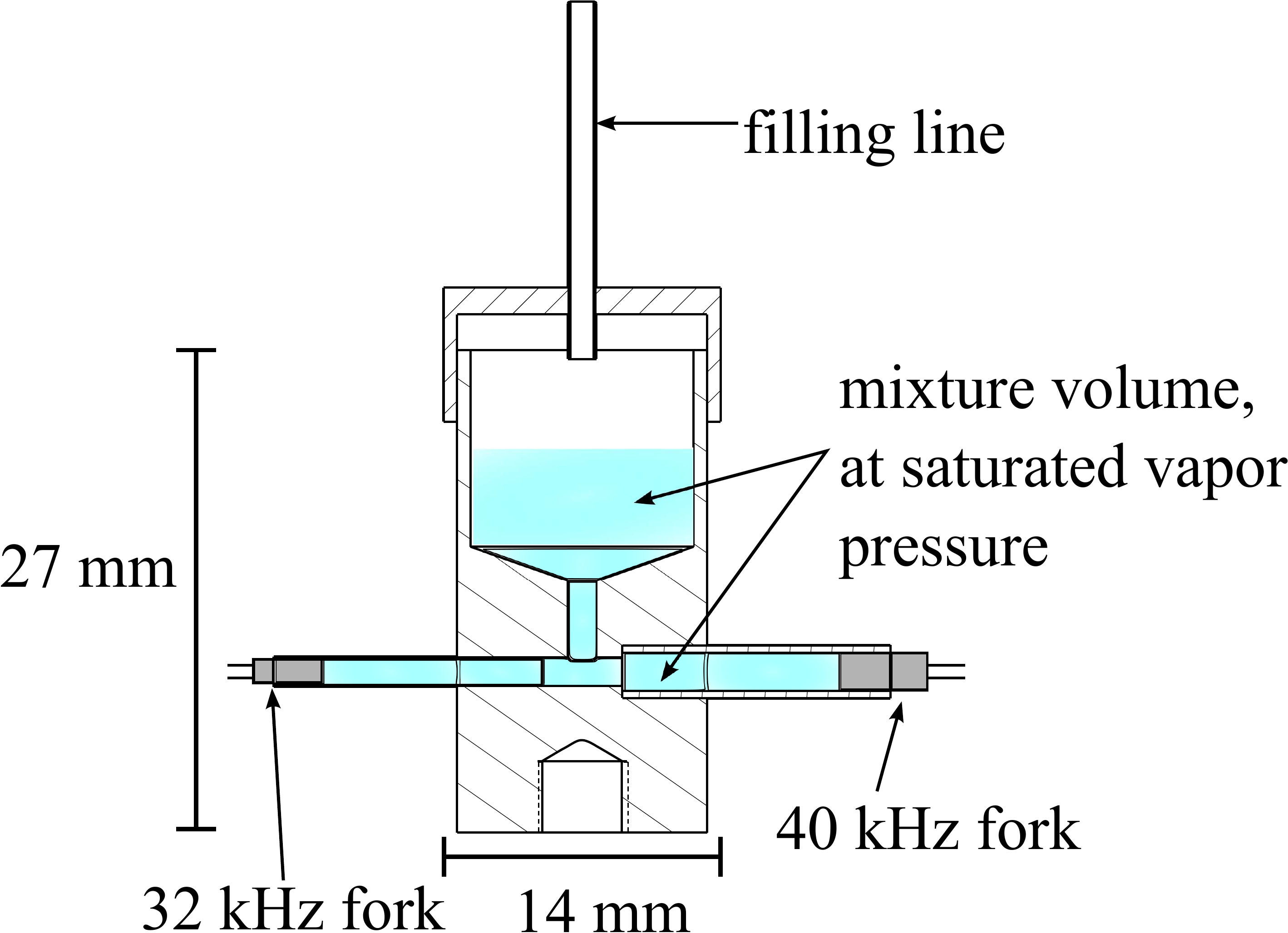}

\caption{Schematic drawing of the \protect\textsuperscript{3}He--\protect\textsuperscript{4}He
mixture cell. The buffer volume above the horizontal quartz tuning
fork volume helps maintain the liquid under saturated vapor pressure.\label{fig:Schematic-cell}}

\end{figure}

Our $2\,\mathrm{cm^{3}}$ experimental cell, shown in Fig. \ref{fig:Schematic-cell},
was a simple copper container that had two horizontal tubes soldered
at the bottom to house two quartz tuning forks. The copper container
itself acted as a buffer volume to ensure that the liquid in the cell
was always under saturated vapor pressure, and that the forks were
always properly immersed in liquid. The cell was installed in a glass
dewar, which could be filled with liquid \textsuperscript{4}He and
then pumped to reach temperatures down to about $1.6\,\mathrm{K}$.
The temperature of the glass dewar was adjusted by combination of
throttling the pumping and a computer controlled heater. Temperature
was monitored with two carbon resistors: one was placed directly on
the cell, while the other was fixed on the support structure. Cell
and bath pressure were measured using \emph{Pfeiffer Vacuum PCR 280}
pressure gauges. The carbon resistors were calibrated against the
vapor pressure during pure \textsuperscript{4}He measurements.

\textsuperscript{3}He -- \textsuperscript{4}He mixtures were prepared
at room temperature. We started with $25\,\mathrm{mmol}$ of commercial
quality pure \textsuperscript{4}He and systematically added mixture
of known concentration to obtain the desired composition, finally
ending up with $94\,\mathrm{mmol}$ of $9\,\%$ molar concentration
mixture. After that, we conducted measurements using mixture taken
directly from $6.0\,\%$, and $11.0\,\%$ storage tanks. We used these
measurements to calibrate our quartz tuning fork to cross check the
concentrations of the earlier used mixtures.

\subsection{Quartz Tuning Fork}

We used \emph{ECS-.327-8-14X} $32.768\,\mathrm{kHz}$ quartz tuning
fork oscillator, which was excited by a function generator, and the
signal was detected by a lock-in amplifier. We also had another fork
with a different resonance frequency and larger physical size installed,
but it behaved erratically, and it then could not be used to produce
any meaningful experimental data. We measured the fork in so-called
tracking mode \cite{Pentti_Rysti_Salmela}, in which a computer program
determines the resonance frequency and the width of the resonance
from a single measurement point close to the actual resonance frequency,
assuming that the shape of the resonance is Lorentzian. The tracking
mode enables us to repeat the measurement every few seconds instead
of minutes, as it would take if we were to record entire resonance
spectra.

Quartz tuning forks do not respond to the temperature of the liquid
directly, rather they sense it through the change in viscosity and
density of the liquid due to temperature. This means that a standing
second sound wave is invisible to the fork, as it cannot cause piezoelectric
response. 

But second sound can create first sound. Such driven first sound has
the same wave characteristics as the second sound wave that is generating
it, but as pressure wave, it can push on the fork causing a piezoelectric
response. When the first sound decouples from second sound, standing
second sound wave is no longer able to drive first sound, and the
anomaly in the fork resonance behavior disappears.

These anomalies, caused by first sound driven by second sound, are
usually simply called second sound resonances, even if it is slightly
inaccurate. The velocity of second sound ranges approximately between
$0\,\mathrm{m/s}$ and $40\,\mathrm{m/s}$, in the temperature and
concentration range of our experiment. Largest values are obtained
at low temperatures and at high concentrations, while near $T_{\lambda}$
it tends to zero, as second sound ceases to exist \cite{Pentti_Rysti_Salmela,Lane_Fairbank,Maurer_Herlin,Klerk_Hudson,King_Fairbank}.
Since the second sound velocity has a significant temperature and
concentration dependence, there exist numerous possible standing sound
wave modes that can be observed with a quartz tuning fork \cite{Tuoriniemi_etal_model}.
\begin{figure*}
\includegraphics[width=17cm]{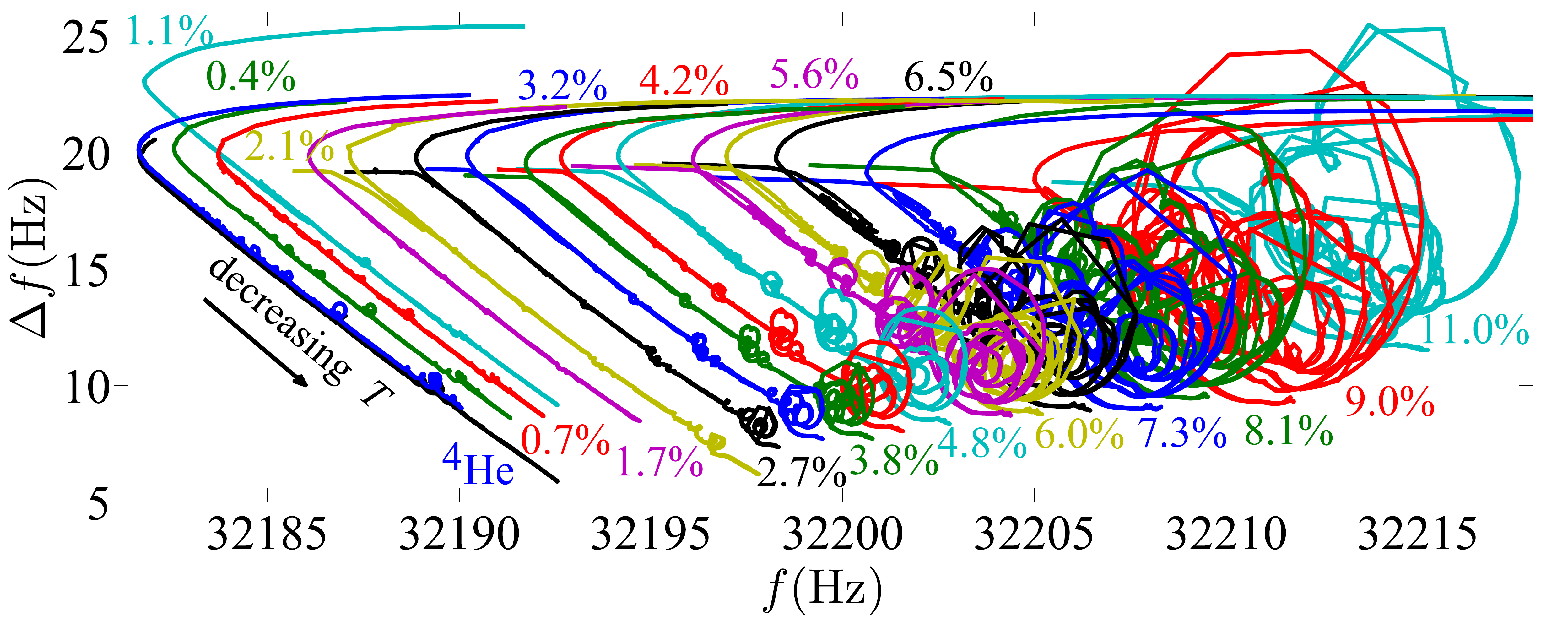}

\caption{Resonance width ($\Delta f$) of the $32\,\mathrm{kHz}$ quartz tuning
fork versus its resonance frequency ($f$) with number of \protect\textsuperscript{3}He
concentrations. The figure contains temperature sweeps both down and
up, falling basically on top of each other. The $\lambda$-point appears
as tilted V-shape near $20\,\mathrm{Hz}$ width, and below that the
linear decline is due to change in temperature. Second sound resonances
appear as loops, whose magnitude is proportional to the coupling strength
between second and first sound. There are two pure \protect\textsuperscript{4}He
measurement sets to demonstrate the reproducibility. Also, note that
the $11.0\,\%$ dataset does not include the lowest temperature resonances
as it was done faster than the other sweeps. Its purpose was to be
a reference point for our \protect\textsuperscript{3}He concentration
analysis.\label{fig:width-freq_ALL}}
\end{figure*}

We do not know the exact shape of the standing wave mode within the
liquid surrounding the fork. This would be rather difficult to determine
as the geometry of the fork is non-trivial. But, the quantity of interest
here is the coupling between first sound and second sound, and this
is independent from the quartz oscillator. The fork geometry only
determines how many second sound resonances we are able to see, \linebreak
at which temperature and concentration they appear, and how strong
an anomaly they create.

The wave mode generated by the quartz tuning fork is likely not just
pure first sound, but rather some combination of first and second
sound. Second sound contribution generated by the fork can couple
back through single sound conversion (second sound to first sound),
and first sound contribution through double sound conversion (first
sound to second sound an back to first). In a single conversion, the
coupling factor is just $\alpha$, but in a double conversion the
amplitude of the resonating first sound mode is proportional to the
product of the coupling factors, $\alpha\beta$. But since only the
coupling factor $\alpha$ has a significant temperature and \textsuperscript{3}He
concentration dependence, it chiefly determines the coupling/decoupling
behavior. The slim temperature dependence of $\beta$ meant that,
we were not able to discern whether the pressure wave affecting the
fork had come about through single sound conversion from second sound
generated by the fork, or through double sound conversion from first
sound generated by the fork. The bottom line is that the determining
factor is the coupling coefficient $\alpha$. When it becomes very
small, in either case, second sound can no longer drive first sound,
and there would then no longer exist a first sound mode that can couple
back to the fork altering its resonance behavior.

\section{Results\label{sec:Results}}

Temperature sweeps were carried out from $\lambda$-point, down to
about $1.6\,\mathrm{K}$ and back. Near $T_{\lambda}$ the sweep rate
had to be quite slow, $0.5\,\mathrm{mK/min}$, since there were many
small second sound resonances there. Below about $2\,\mathrm{K}$,
we could increase the rate to $1.5\,\mathrm{mK/min}$ as the resonances
became more infrequent, and wider in temperature. Fig. \ref{fig:width-freq_ALL}
shows the resonance frequency of the $32\,\mathrm{kHz}$ quartz tuning
fork versus resonance width at different \textsuperscript{3}He concentrations.
In this presentation, anomalies caused by second sound appear as loops.
The second sound resonances appeared on same temperature independent
of the direction of the temperature sweep, to better than $\pm0.6\,\mathrm{mK}$.
Pure \textsuperscript{4}He was measured multiple times, both before
and after the mixture measurements, from which we could estimate the
reproducibility to be about $\pm3\,\mathrm{mK}$. This spread is due
to uncertainty in our temperature determination with the carbon resistors,
rather than any variation in the sound properties. Starting from $2.1\,\%$
\textsuperscript{3}He concentration, there appears a horizontal feature
near the $\lambda$-point, which is caused by the mixing of the helium
isotopes. They do not mix properly until near $T_{\lambda}$, after
which the fork resonance frequency changes rapidly to a new value.

The $1.1\,\%$ dataset is shifted with respect to others due to some
unknown unreproducible phenomenon, possibly related to some impurities
sticking to the fork. We reached this conclusion since the problem
did not repeat itself after we had warmed our cooling system back
to room temperature between measurements. These data can be made compatible
with the others by simple shift, bringing the kink at the $\lambda$-point
to the appropriate position. We emphasize that Fig. \ref{fig:width-freq_ALL}
displays the raw data as measured, with no adjustment or post processing.
\begin{figure}
\subfloat{\includegraphics[clip,width=8cm]{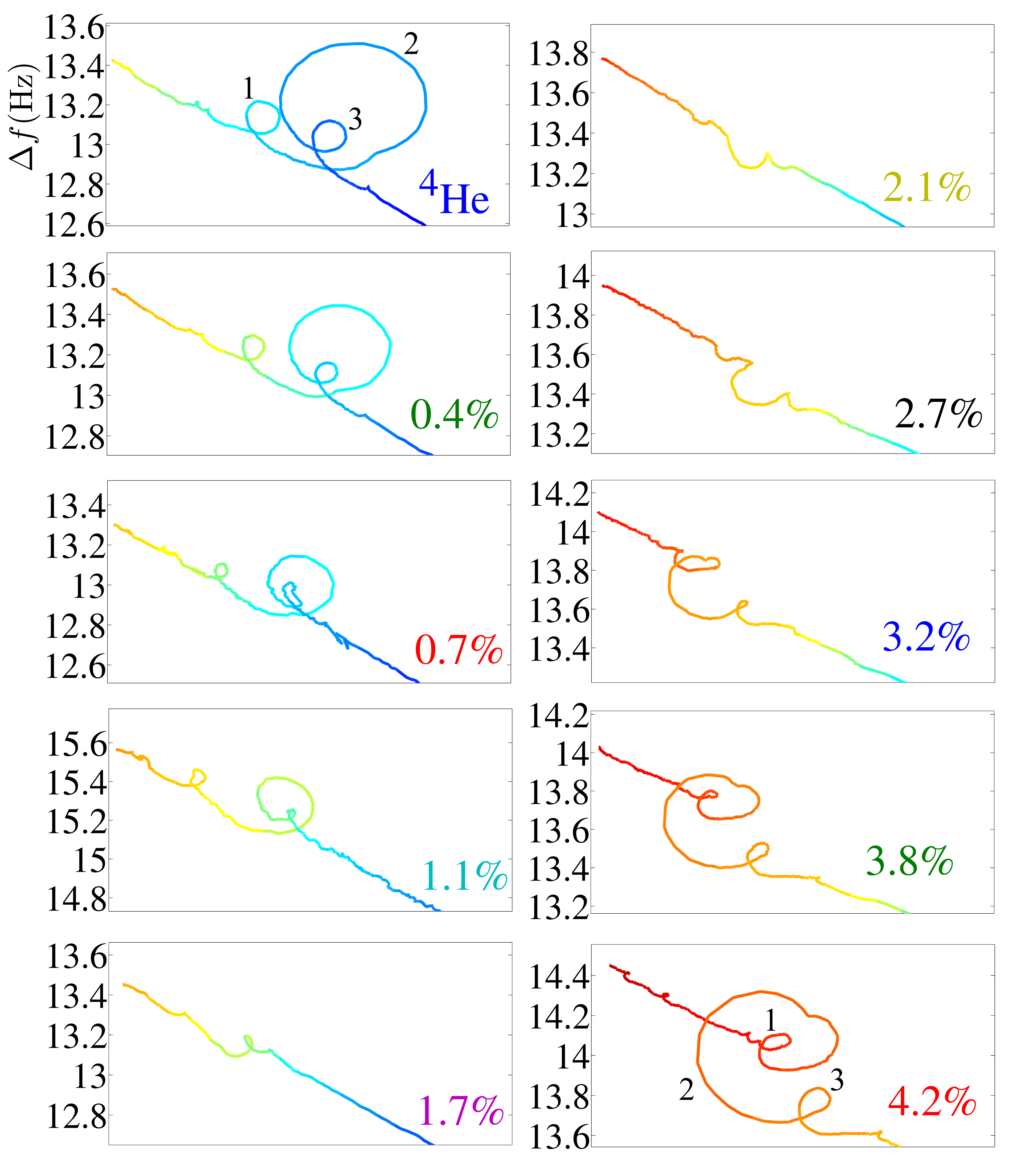}}\vspace{0.01cm}
\center\subfloat{\includegraphics[width=5cm]{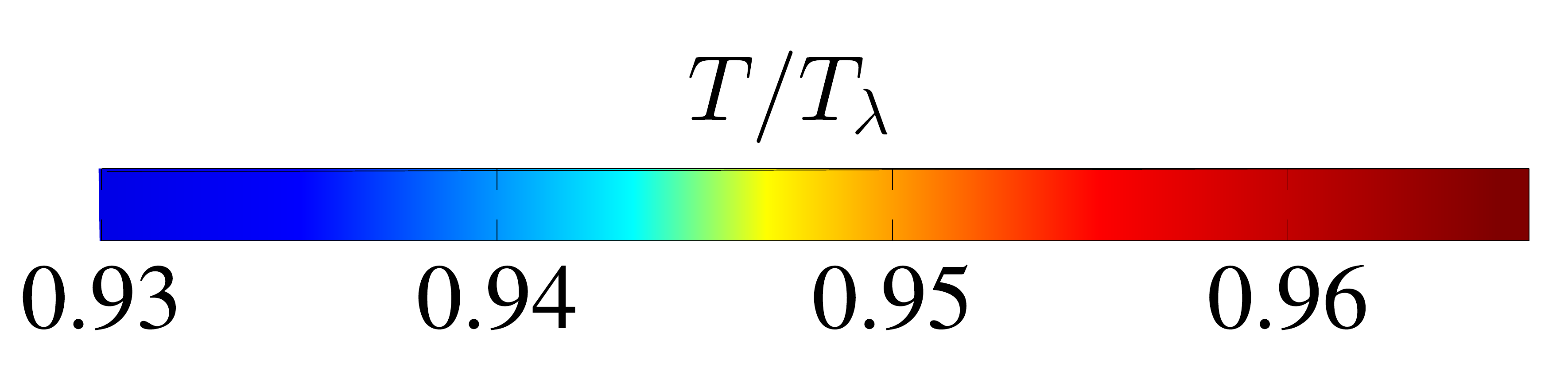}}

\caption{Closer view of the largest high temperature second sound resonances
of Fig. \ref{fig:width-freq_ALL}, followed through the decoupling
region. The color of the line changes according to temperature, showing
that the resonances move to a higher relative temperature as the $^{3}\mathrm{He}$
concentration increases. Even though the shape of the resonances changes,
we can still identify the three biggest resonances of pure $^{4}\mathrm{He}$
also in $4.2\,\%$ mixture. \label{fig:freq-width_ZOOM}}
\end{figure}

Since the sound mode coupling in pure \textsuperscript{4}He is caused
only by the very small thermal expansion, the magnitude of the loops
is also quite small. Remarkably, they become even smaller as some
amount of \textsuperscript{3}He is added. The addition of \textsuperscript{3}He
initially weakens the coupling between the sound modes, as was predicted
by our calculations of the coupling factor $\alpha$ in Section \ref{sec:Sound-Conversion}.
The second sound anomalies vanish somewhere between $1.1\,\%$ and
$2.1\,\mathrm{\%}$ concentrations, which is indicative of the decoupling
between the two sound modes. As the \textsuperscript{3}He concentration
is further increased, the second sound resonances reappear, eventually
becoming significantly larger than they were in pure \textsuperscript{4}He,
because now the \textsuperscript{3}He contribution to the coupling
is dominant.

What is more, at $2.1\,\%$ concentration, there appears a new set
of second sound resonances at the low temperature end of the sweep,
which were absent in pure \textsuperscript{4}He. Temperature sweeps
between pure \textsuperscript{4}He and $2.1\,\%$ \textsuperscript{3}He
concentration, except for $1.1\,\%$ measurement, were not extended
down to the lowest reachable temperature, since we had not initially
expected to find anything there. Since the $1.1\,\%$ sweep continued
to a lower temperature than the ones next to it, we can conclude that
these resonances had not yet appeared at this concentration.

\textsuperscript{3}He concentrations shown in Fig. \ref{fig:width-freq_ALL}
were obtained by making a linear fit to the background decline of
each temperature sweep, and using $6.0\,\%$, and $11.0\,\%$ concentrations
as reference points, since their mixture was taken directly from room
temperature storage tanks with known concentrations. The uncertainty
of all concentrations was estimated to be $\pm0.3$ percentage points.
The concentration values obtained from the tuning fork analysis were
roughly 0.5 percentage points less than the concentration estimated
while preparing the gas mixture at room temperature.

In Fig. \ref{fig:freq-width_ZOOM}, we take a closer look of one set
of second sound resonances illustrating their behavior near the decoupling
region. Even though their shape changes as the \textsuperscript{3}He
concentration is increased, we can still identify the correspondent
second sound resonances because they always appear in the same sequence
--- a larger resonance flanked by two smaller resonances, plus a number
of tinier ones, in our example. When the coupling is at its weakest,
only the large resonance remains barely visible, and it too would
seem to disappear somewhere between $1.7\,\%$ and $2.1\,\%$ concentrations.
Even if the fork resonance width of the $1.1\,\%$ measurement set
is in slightly different range than the others, when also considering
temperature, that dataset fits in quite well with the others.
\begin{figure}
\includegraphics[width=9cm]{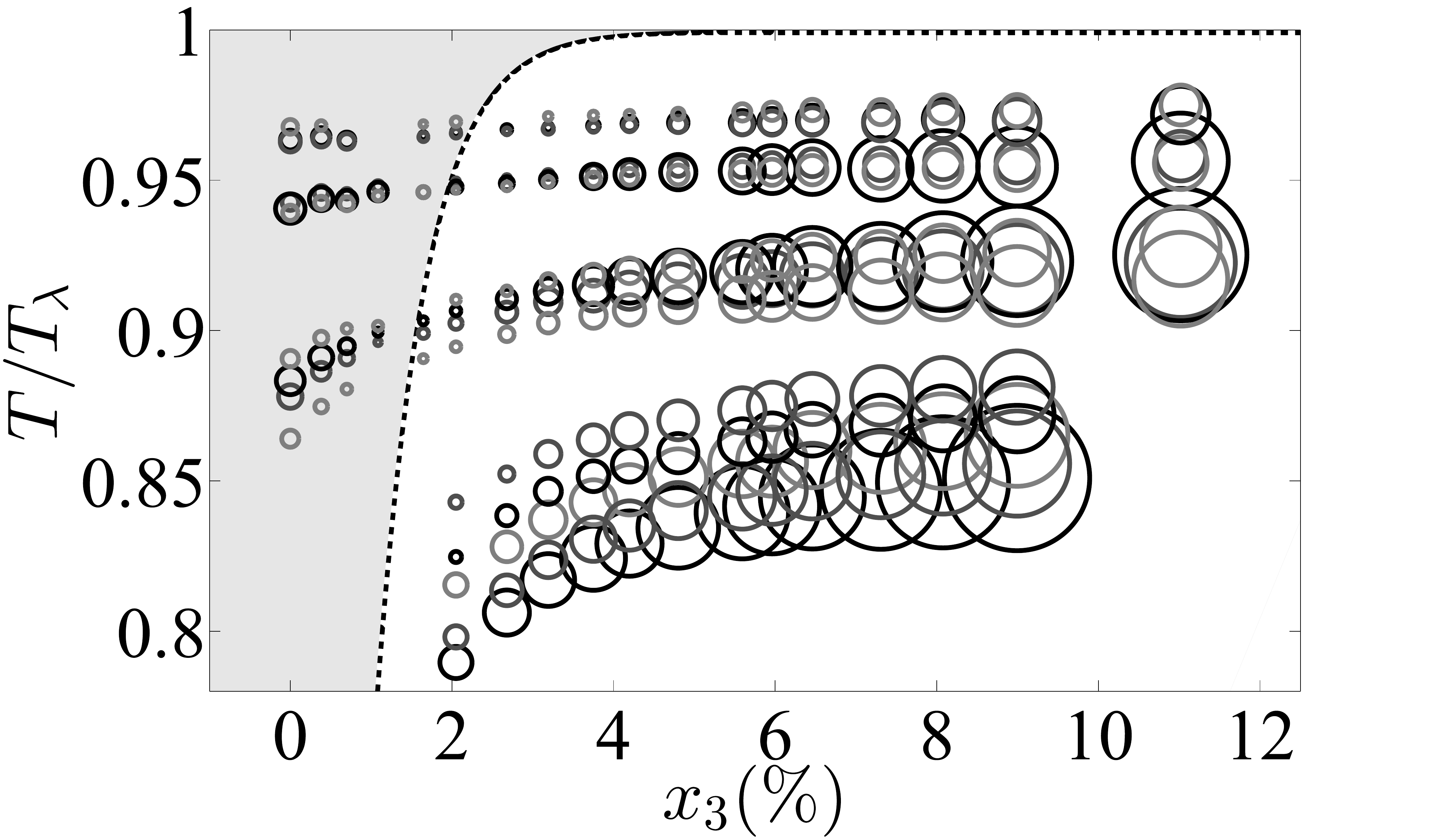}

\caption{Locations of the second sound resonances followed through the decoupling
region in $^{3}\mathrm{He}$ concentration -- relative temperature
plane, as well as their amplitude (represented by the size of the
circle). Dashed line corresponds the dashed line of Fig. \ref{fig:decoupling_meas},
which separates the second sound resonances before the decoupling
from those after the decoupling.\label{fig:Circles}}
\end{figure}

We determined the amplitude for each second sound resonance as the
maximum deviation from the fork's background slope, and they were
then normalized to the value in the $9.0\,\%$ measurement set. The
temperature of each resonance was defined to be the point of the maximum
deviation. These are shown in Fig. \ref{fig:Circles}. Not all second
sound resonances of Fig. \ref{fig:width-freq_ALL} are included, but
rather the selected few that we were able to follow through the decoupling
region with sufficient certainty. At high temperatures, or at high
concentrations, the anomalies appear almost at the same relative temperature,
but otherwise they start to bend to lower temperatures.

\begin{figure}
\includegraphics[width=9cm]{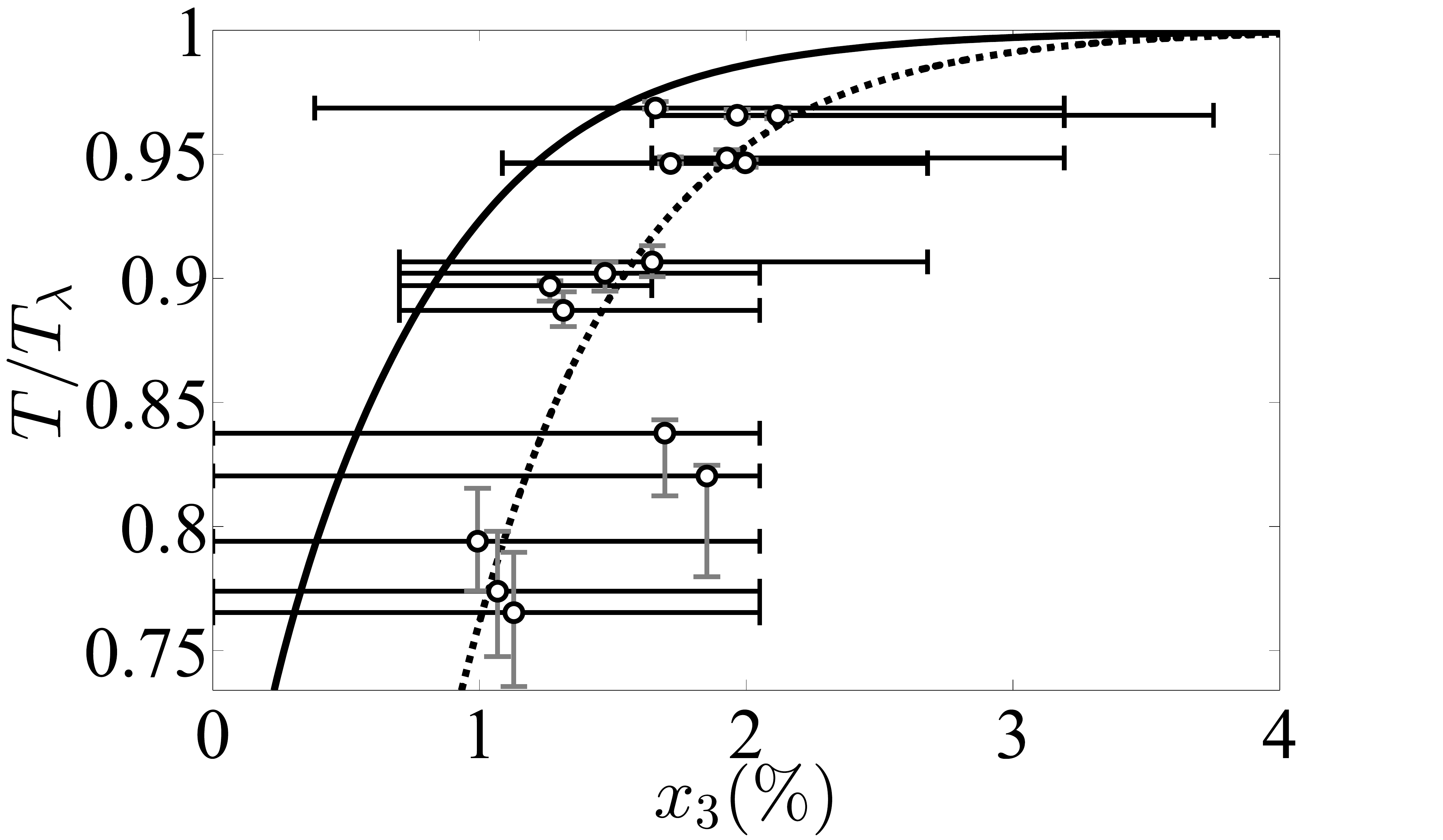}

\caption{Decoupling temperatures and concentrations interpolated from experimental
data compared to the calculated data shown by the solid line (\emph{cf.}
Fig. \ref{fig:decoupling_calc}). Dashed line is guide to the eye
drawn through the evaluated decoupling points. \label{fig:decoupling_meas}}
\end{figure}

The amplitude data allows us to interpolate the locations where the
second sound resonances would disappear, as the sound modes become
decoupled. These are shown in Fig. \ref{fig:decoupling_meas}, where
they are compared against the decoupling behavior calculated in Section
\ref{sec:Sound-Conversion}. The errorbars were determined from concentrations
and temperatures where the anomalies had definitely not yet disappeared,
or had definitely appeared again. We also extrapolated the lowest
temperature second sound resonance data to find where they would disappear,
and since they were not visible in pure \textsuperscript{4}He, their
errorbars extend all the way to zero concentration. Decoupling points
determined from our measurements lie systematically at higher \textsuperscript{3}He
concentrations than the calculated values. This would suggest that
either the \textsuperscript{3}He contribution to the coupling is
slightly smaller than the calculations indicated, or that there exists
some additional coupling contribution that we had not taken into account.

\section{Conclusions\label{sec:Conclusions}}

We studied coupling between first sound and second sound in \textsuperscript{3}He
-- superfluid \textsuperscript{4}He mixtures, down to $1.6\,\mathrm{K}$
temperature under saturated vapor pressure. Velocity of second sound
is such that it can form standing waves around a quartz tuning fork
immersed in superfluid. Second sound drives first sound with the same
geometry, and this first sound perturbation can be detected by the
fork as an anomalous resonance behavior. Since the specific second
sound resonances always appear under the same conditions due to the
nature of standing waves, they can be used, for example, to indicate
fixed points of temperature with good accuracy \cite{Salmela_fixedPoints}.

We confirmed, that at certain concentrations and temperatures, these
second sound anomalies briefly disappear, before reappearing as the
\textsuperscript{3}He concentration is increased. This behavior is
a result of the competing contributions to the coupling between the
two sound modes. When the sound modes become decoupled, the standing
second sound wave can still exist, but it can no longer create first
sound, and hence it becomes invisible to the quartz tuning fork. Our
calculations, that revised the results presented in an earlier publication
\cite{Brusov}, predicted exactly this kind of behavior, but they
projected the decoupling to occur at somewhat lower \textsuperscript{3}He
concentration.
\begin{acknowledgments}
We thank J. Rysti for valuable discussions. This work was supported
by the Academy of Finland CoE 20122017, Grant No. 250280 LTQ. We also
acknowledge the provision of facilities by Aalto University at OtaNano
- Low Temperature Laboratory.
\end{acknowledgments}

\bibliographystyle{apsrev4-1}
\bibliography{decoupling}

\end{document}